\documentclass[aps,prl,reprint,groupedaddress]{revtex4-1}
\usepackage{graphicx} 
\usepackage{dcolumn} 
\usepackage{bm}
\usepackage{hyperref}

\begin{document}

\title{Production of heavy isotopes in transfer reactions by collisions of $^{238}$U+$^{238}$U}

\author{Zhao-Qing Feng}
\email{fengzhq@impcas.ac.cn}
\author{Gen-Ming Jin}
\author{Jun-Qing Li}
\affiliation{Institute of Modern Physics, Chinese Academy of
Sciences, Lanzhou 730000, People's Republic of China}

\date{\today}

\begin{abstract}
The dynamics of transfer reactions in collisions of two very heavy
nuclei $^{238}$U+$^{238}$U is studied within the dinuclear system
(DNS) model. Collisions of two actinide nuclei form a super heavy
composite system during a very short time, in which a large number
of charge and mass transfers may take place. Such reactions have
been investigated experimentally as an alternative way for the
production of heavy and superheavy nuclei. The role of collision
orientation in the production cross sections of heavy nuclides is
analyzed systematically. Calculations show that the cross sections
decrease drastically with increasing the charged numbers of heavy
fragments. The transfer mechanism is favorable to synthesize heavy
neutron-rich isotopes, such as nuclei around the subclosure at N=162
from No (Z=102) to Db (Z=105).
\begin{description}
\item[PACS number(s)]
25.70.Jj, 24.10.-i, 25.60.Pj
\end{description}
\end{abstract}

\maketitle

The synthesis of superheavy nuclei (SHN) is motivated with respect
to searching the "island of stability", which is predicted
theoretically, and has resulted in a number of experimental studies.
There are mainly two sorts of reaction mechanism to produce heavy
and superheavy nuclei, namely, multi-nucleon transfer reactions in
collisions of two actinide nuclei \cite{Hu77,Sc78} and
fusion-evaporation reactions with a neutron-rich nuclide bombarding
a heavy nucleus near shell closure, such as the cold fusion
reactions at GSI (Darmstadt, Germany) \cite{Ho00} and the $^{48}$Ca
bombarding the actinide nuclei at FLNR in Dubna (Russia)
\cite{Og07}. However, the decay chains of the nuclei formed by the
hot fusion reactions are all neutron-rich nuclides and do not
populate presently the known nuclei. Meanwhile, the superheavy
isotopes synthesized by the cold fusion and the $^{48}$Ca induced
reactions are all far away from the doubly magic shell closure
beyond $^{208}$Pb at the position of protons Z=114-126 and neutrons
N=184. Multi-nucleon transfer reactions in collisions of two
actinides might be used to fill the region and examine the influence
of the shell effect in the production of heavy isotopes. The
production of neutron-rich heavy or superheavy nuclei in low-energy
collisions of actinide nuclei was proposed initially by Zagrebaev
\emph{et al.} based on the assumption that the shell effects
continue to play a significant role in multi-nucleon transfer
reactions \cite{Za06}.

The cross sections of the heavy fragments in strongly damped
collisions between very heavy nuclei were found to decrease very
rapidly with increasing the atomic number \cite{Hu77,Sc78}.
Calculations by Zagrebaev and Greiner with a model based on
multi-dimensional Langevin equations \cite{Za07} showed that the
production of the survived heavy fragments with the charged number
Z$>$106 is rare because of the very small cross sections at the
level of 1 pb and even below 1 pb. However, neutron-rich isotopes of
Fm and Md were produced at the larger cross section of 0.1 $\mu$b.
The evolution of the composite system in the damped collisions is
mainly influenced by the incident energy and the collision
orientation. Recently, the time dependent Hartree-Fock (TDHF)
approach \cite{Go09} and improved quantum molecular dynamics (ImQMD)
model \cite{Ti08} were also used to investigate the dynamics in
collisions of $^{238}$U+$^{238}$U.

In this work, we use a dinuclear system (DNS) model to investigate
the dynamics of the damped collisions of two very heavy nuclei, in
which the nucleon transfer is coupled to the relative motion by
solving a set of microscopically derived master equations by
distinguishing protons and neutrons \cite{Fe06,Fe07}. In order to
treat the diffusion process along proton and neutron degrees of
freedom in the damped collisions, the distribution probability is
obtained by solving a set of master equations numerically in the
potential energy surface of the DNS. The time evolution of the
distribution probability $P(Z_{1},N_{1},E_{1},t)$ for fragment 1
with proton number $Z_{1}$ and neutron number $N_{1}$ and with
excitation energy $E_{1}$ is described by the following master
equations,
\begin{widetext}
\begin{eqnarray}
\frac{d P(Z_{1},N_{1},E_{1},t)}{dt}=&&\sum_{Z_{1}^{\prime
}}W_{Z_{1},N_{1};Z_{1}^{\prime},N_{1}}(t)\left[
d_{Z_{1},N_{1}}P(Z_{1}^{\prime},N_{1},E_{1}^{\prime},t)-d_{Z_{1}^{\prime
},N_{1}}P(Z_{1},N_{1},E_{1},t)\right]+\sum_{N_{1}^{\prime
}}W_{Z_{1},N_{1};Z_{1},N_{1}^{\prime}}(t) \nonumber \\
&& \left[
d_{Z_{1},N_{1}}P(Z_{1},N_{1}^{\prime},E_{1}^{\prime},t)-d_{Z_{1},N_{1}^{\prime}}P(Z_{1},N_{1},E_{1},t)\right]-
\left[\Lambda_{qf}(\Theta(t))+\Lambda_{fis}(\Theta(t))
\right]P(Z_{1},N_{1},E_{1},t).
\end{eqnarray}
\end{widetext}
Here the $W_{Z_{1},N_{1};Z_{1}^{\prime},N_{1}}$
($W_{Z_{1},N_{1};Z_{1},N_{1}^{\prime}}$) is the mean transition
probability from the channel $(Z_{1},N_{1},E_{1})$ to
$(Z_{1}^{\prime},N_{1},E_{1}^{\prime})$ (or $(Z_{1},N_{1},E_{1})$ to
$(Z_{1},N_{1}^{\prime},E_{1}^{\prime})$), and $d_{Z_{1},N_{1}}$
denotes the microscopic dimension corresponding to the macroscopic
state $(Z_{1},N_{1},E_{1})$. The sum is taken over all possible
proton and neutron numbers that fragment
$Z_{1}^{\prime},N_{1}^{\prime}$ may take, but only one nucleon
transfer is considered in the model with the relation $Z_{1}^{\prime
}=Z_{1}\pm 1$ and $N_{1}^{\prime }=N_{1}\pm 1$. The excitation
energy $E_{1}$ is determined by the dissipation energy from the
relative motion and the potential energy surface of the DNS. The
motion of nucleons in the interacting potential is governed by the
single-particle Hamiltonian \cite{Fe06,Fe07}. The evolution of the
DNS along the variable $R$ leads to the quasifission of the DNS. The
quasifission rate $\Lambda_{qf}$ and the fission rate
$\Lambda_{fis}$ of the heavy fragment are estimated with the
one-dimensional Kramers formula. The local temperature is given by
the Fermi-gas expression $\Theta=\sqrt{\varepsilon^{\star}/a}$
corresponding to the local excitation energy $\varepsilon^{\star}$
and the level density parameter $a=A/12$ MeV$^{-1}$ \cite{Fe07}.

\begin{figure*}
\includegraphics{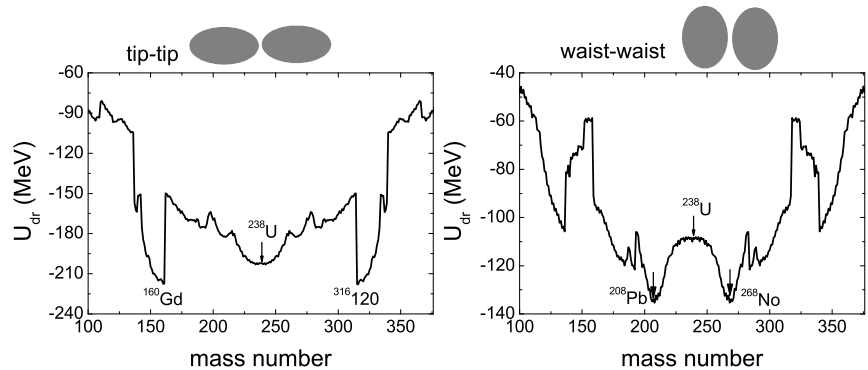}
\caption{\label{fig:wide} Driving potentials of the tip-tip and
waist-waist collisions in the reaction $^{238}$U+$^{238}$U.}
\end{figure*}

In the relaxation process of the relative motion, the DNS will be
excited by the dissipation of the relative kinetic energy. The local
excitation energy is determined by the excitation energy of the
composite system and the potential energy surface of the DNS. The
potential energy surface (PES) of the DNS is given by
\begin{eqnarray}
U(\{\alpha\})=&&B(Z_{1},N_{1})+B(Z_{2},N_{2})-
\left[B(Z,N)+V^{CN}_{rot}(J)\right] \nonumber\\
&&+V(\{\alpha\})
\end{eqnarray}
with $Z_{1}+Z_{2}=Z$ and $N_{1}+N_{2}=N$ \cite{Fe09}. Here the
symbol $\{\alpha\}$ denotes the sign of the quantities $Z_{1},
N_{1}, Z_{2}, N_{2}; J, R; \beta_{1}, \beta_{2}, \theta_{1},
\theta_{2}$. The $B(Z_{i},N_{i}) (i=1,2)$ and $B(Z,N)$ are the
negative binding energies of the fragment $(Z_{i},N_{i})$ and the
compound nucleus $(Z,N)$, respectively, which are calculated from
the liquid drop model, in which the shell and the pairing
corrections are included reasonably. The $V^{CN}_{rot}$ is the
rotation energy of the compound nucleus. The $\beta_{i}$ represent
the quadrupole deformations of the two fragments. The $\theta_{i}$
denote the angles between the collision orientations and the
symmetry axes of deformed nuclei. The interaction potential between
fragment $(Z_{1},N_{1})$ and $(Z_{2},N_{2})$ includes the nuclear,
Coulomb and centrifugal parts; the detailed calculations are given
in Ref. \cite{Fe07}. In the calculation, the distance $R$ between
the centers of the two fragments is chosen to be the value at the
touching configuration, in which the DNS is assumed to be formed. So
the PES depends on the proton and neutron numbers of the fragments.
Shown in Fig. 1 is the calculated PES as functions of mass numbers
of fragments for the two cases of the nose-nose and side-side
orientations. Dissipation to heavier fragment by nucleon transfer is
hindered in the nose-nose collisions and a pocket appears around the
nucleus $^{316}$120. The side-side orientation is easily to reach
the subclosure $^{268}$No, but is also hindered if further
dissipating the high-mass region.

\begin{figure*}
\includegraphics{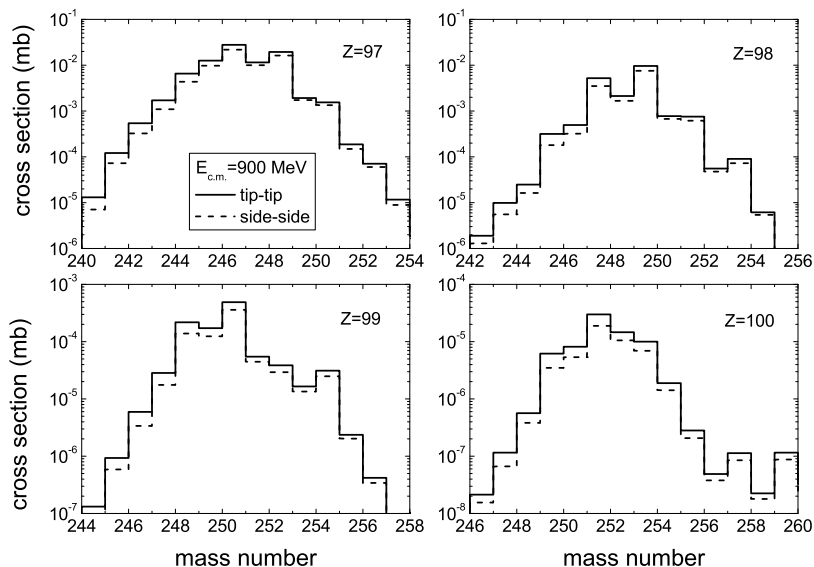}
\caption{\label{fig:wide} Comparison of the production cross
sections of the Bk, Cf, Es and Fm isotopes in collisions of
$^{238}$U with $^{238}$U for the tip-tip and waist-waist
orientations, respectively.}
\end{figure*}

\begin{figure}
\includegraphics[width=8 cm]{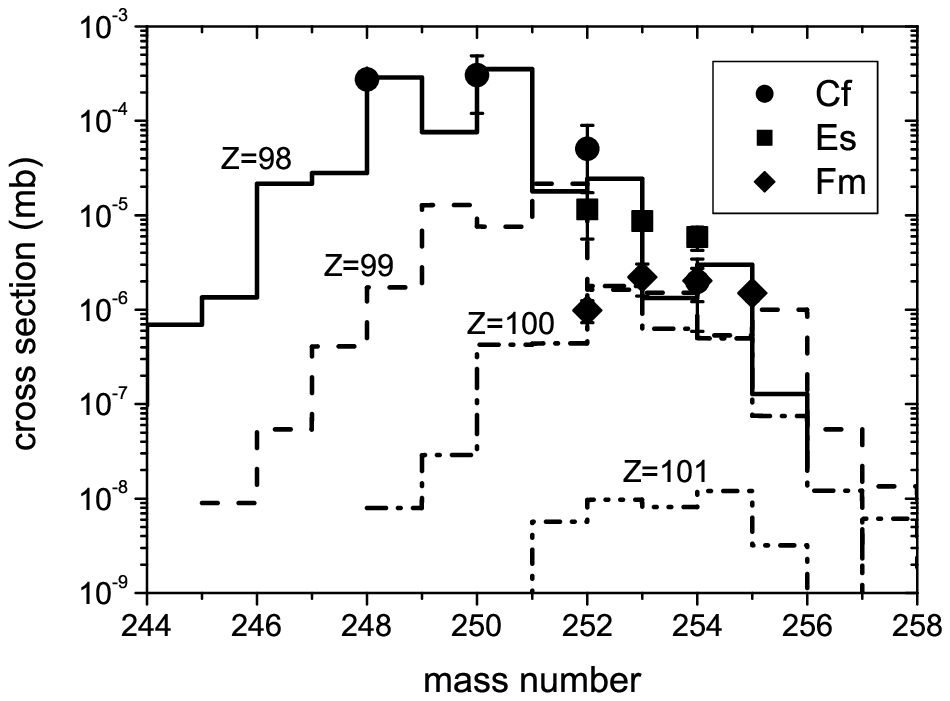}
\caption{\label{fig:epsart} Calculated mass distributions of the Cf,
Es, Fm and Md isotopes at E$_{c.m.}$=800 MeV and compared with the
available experimental data \cite{Sc78}.}
\end{figure}

The cross sections of the primary fragments (Z$_{1}$, N$_{1}$) after
the DNS reaches the relaxation balance are calculated as follows:
\begin{equation}
\sigma_{pr}(Z_{1},N_{1})=\frac{\pi \hbar^{2}}{2\mu E_{c.m.}}
\sum_{J=0}^{J_{\max}}(2J+1)P(Z_{1},N_{1},\tau_{int}).
\end{equation}
The interaction time $\tau _{int}$ in the dissipation process of two
colliding partners is dependent on the incident energy $E_{c.m.}$ in
the center-of-mass (c.m.) frame. and the angular momentum $J$, which
is calculated by using the deflection function method \cite{Li83}
and has the value of few 10$^{-20}$s. The survived fragments are the
decay products of the primary fragments after emitting the particles
and $\gamma$ rays in competition with fission. The cross sections of
the survived fragments are given by
\begin{eqnarray}
\sigma_{sur}(Z_{1},N_{1})=&&\frac{\pi \hbar^{2}}{2\mu E_{c.m.}}
\sum_{J=0}^{J_{\max}}(2J+1)P(Z_{1},N_{1},E_{1},\tau_{int}) \nonumber\\
&& \times W_{sur}(E_{1},xn,J),
\end{eqnarray}
where the $E_{1}$ is the excitation energy of the fragment (Z$_{1}$,
N$_{1}$). The maximal angular momentum is taken as $J_{\max}=200$
that includes all partial waves in which the transfer reactions may
take place. The survival probability $W_{sur}$ of each fragment can
be estimated by using the statistical approach \cite{Fe07}.

Dynamics of the damped collisions was investigated by Zagrebaev and
Greiner in detail with a model based on multi-dimensional Langevin
equations \cite{Za07,Za08}. Larger cross sections in the production
of neutron-rich heavy isotopes in collisions of two actinides were
pointed out. Within the framework of the DNS model, we calculated
the production of the survived fragments in collisions of
$^{238}$U+$^{238}$U for the nose-nose and side-side orientations at
900 MeV center-of-mass energy as shown in Fig. 2. The collisions of
the side-side case need to overcome the higher barrier of the
interaction potential in the formation of the DNS. But it is
favorable to transfer nucleon by the master equations in the driving
potential and form the target-like fragments. Situation is opposite
for the nose-nose collisions. So both cases give a similar result in
the production of the Bk, Cf, Es and Fm isotopes. Comparison of the
calculated mass distributions and the experimental data of the
survived fragments at E$_{c.m.}$=800 MeV is shown in Fig. 3. The
cross sections decrease drastically with the atomic numbers of
fragments. The calculated results are the case of the nose-nose
collisions, which have the height of the interaction potential at
the touching configuration with the value 713 MeV. In the collisions
of such heavy systems, there is no Coulomb barrier in the
approaching process of two colliding partners. A number of nucleon
transfers take place in the reactions of two actinides owing to the
dynamical deformations and the fluctuations of all collective
degrees of freedoms in the model of Zagrebaev and Greiner
\cite{Za07}. We assumed the DNS is formed at the touching
configuration in the collisions of two very heavy nuclei. The
nucleon transfer is governed by the driving potential in competition
with the quasifission of the DNS because of the collision dynamics.
The larger quasifission rate of such systems results in the DNS
quickly decays into two fragments because there is no potential
pockets. The inner excitation energy of the DNS is dissipated from
the kinetic energy of the relative motion overcoming the height of
the interaction potential of two colliding nuclei at the touching
configuration. The value of the side-side orientation is 813.5 MeV.
Inclusion of all orientations in the low-energy damped collisions is
important for correctly estimating the cross sections of the primary
and survived fragments.

\begin{figure*}
\includegraphics{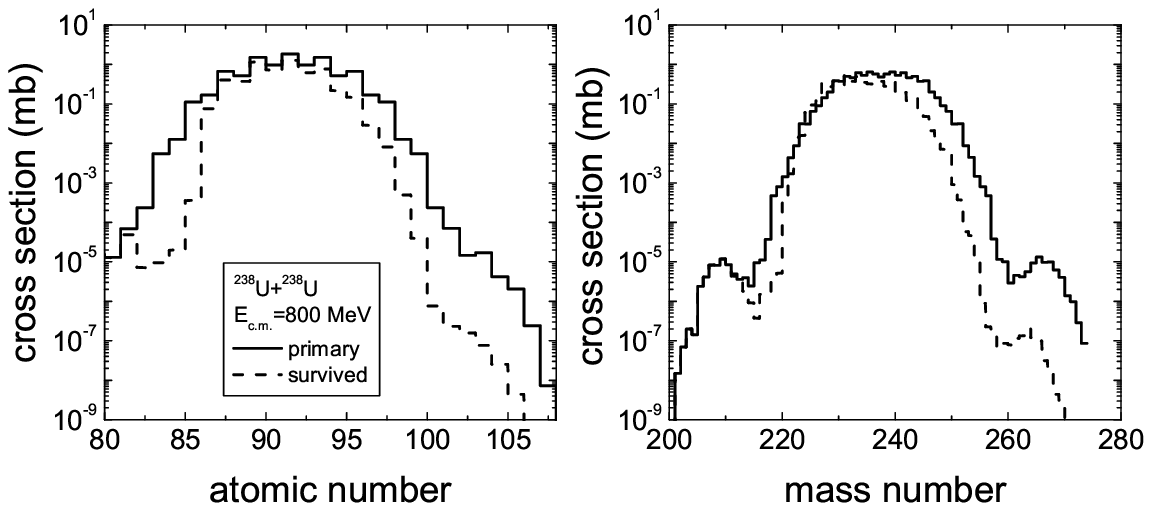}
\caption{\label{fig:wide} Cross sections as functions of the charged
and mass numbers of the primary and survived fragments at
E$_{c.m.}$=800 MeV, respectively.}
\end{figure*}

Shown in Fig. 4 is the calculated cross sections of the primary and
survived fragments by using the Eqs (3) and (4) as functions of the
charged numbers and mass numbers at the incident energy
E$_{c.m.}$=800 MeV, respectively. In the damped collisions, the
primary fragments result from a number of nucleon transfer in the
relaxation process of the colliding partners. The giant composite
system retains a very short time of several tens 10$^{-22}$ s due to
the strong Coulomb repulsion. Calculations from the TDHF method
showed that the collision time depended on the orientation of the
colliding system \cite{Go09}. The cross sections in the production
of heavy target-like fragments (Z$>$92) decrease drastically with
the atomic numbers of the fragments. Therefore, the mechanism of the
low-energy transfer reactions in collisions of two very heavy nuclei
is not suitable to synthesize superheavy nuclei (Z$>$106) because of
the smaller cross sections at the level of 1 pb and even below 1 pb.
Similar results were also obtained in Ref. \cite{Za07}. However, the
production of the survived fragments around the subclosure N=162 has
a larger cross section. Calculated cross sections as functions of
the mass numbers appear a bump near the isotopes of the subclosure.
Experimental works for studying the influence of the shell closure
in the production of the neutron-rich isotopes should be performed
in the near future. It is also a good technique to fill the gap of
the new isotopes between the cold fusion and the $^{48}$Ca induced
reactions.

In summary, the dynamics in collisions of the very heavy system
$^{238}$U+$^{238}$U is investigated within the framework of the DNS
model. The influence of the collision orientations on the production
cross sections of heavy isotopes is discussed systematically. The
low-energy transfer reactions in the damped collisions of the
actinide nuclei are a good mechanism to produce the neutron-rich
heavy isotopes, in which the shell closure plays an important role
in the estimation of the cross sections.

\begin{acknowledgments}
We would like to thank Prof. Werner Scheid for carefully reading the
manuscript. This work was supported by the National Natural Science
Foundation of China under Grants. 10805061, 10775061 and 10975064,
the Special Foundation of the President Fund, the West Doctoral
Project of Chinese Academy of Sciences, and the Major State Basic
Research Development Program under Grant 2007CB815000.
\end{acknowledgments}

\end{document}